\documentclass[12pt]{article}

\begin{document}
\title{Quasiperiodic waves at the onset of zero Prandtl number convection with rotation}
\author{Krishna Kumar$^1$, Sanjay Chaudhuri$^2$, and Alaka Das$^1$\\
$^1$Physics and Applied Mathematics Unit\\ $^2$Statistics and Mathematics Unit\\
Indian Statistical Institute\\ 203, Barrackpore Trunk Road, Calcutta-700~035, India}
\date{12 April, 2001}
\maketitle
\begin{abstract}
 We show the possibility of quasiperiodic waves at the onset
 of thermal convection in a thin horizontal layer of  slowly rotating
 zero-Prandtl number Boussinesq fluid confined between stress-free
 conducting  boundaries. Two independent frequencies emerge due to
 an interaction between a stationary instability and a self-tuned wavy
 instability in presence of coriolis force, if Taylor number is raised
 above a critical value. Constructing a dynamical system for the 
 hydrodynamical problem, the competition  between the interacting 
 instabilities is analyzed. The forward bifurcation from the conductive
 state is self-tuned.\\
 {PACS number(s):47.20.Ky, 47.27.Te}
\end{abstract}

\newpage
Thermal convection in Boussinesq fluids in the  limit
of vanishing  Prandtl number $P (=\nu/\kappa)$~(\cite{spiegel}-\cite{kft})
is of interest for
astrophysical problems ($P \approx 10^{-8}$)  as well as for
liquid metals ($P \approx 10^{-2} - 10^{-3}$). The theoretical
study, in particular with {\it stress-free} boundary conditions,
in the limit of large thermal diffusitivity $\kappa$ has been
considered subtle for a long time because the linearly unstable
two-dimensional (2D) rolls become exact nonlinear solution.
The nonlinearity $ {\bf v\cdot \bf \nabla } \theta $
due to the advection of temperature
fluctuation $\theta$ by the velocity field ${\bf v}$ might be
negligible~\cite{spiegel}. The  nonlinearity ${\bf v \cdot \nabla
v}$, due to the self interaction of velocity field, does not
contribute to saturation for straight (2D) rolls. This led to the
speculation that the zero $P$ limit might involve a singular limit
problem similar to the one with infinite Reynold number in
incompressible fluid dynamics. However, the recent
three-dimensional (3D) direct numerical simulation (DNS) of
zero-Prandtl number Boussinesq equations by Thual~\cite{thual}
showed the saturation instead of indefinite growth of the
solution even with {\it stress-free} boundary conditions. He also
compared the results of zero P equations with that of the full
Oberbeck-Boussinesq equations in the asymptotic limit of vanishing
P, and found complete agreement in two cases.  The saturation of
growing 2D rolls at the onset of convection occurs by generation
of {\it self-tuned} 3D waves, the mechanism of which
was explained in a simple model~\cite{kft}. The results of this
model, in its validity range, agreed well with that of DNS just above
the convective instability. The new nonlocal instability at the onset
occurs purely due to nonlinear effects, while linear equations predict
the stationary instability~\cite{chandra}.

We  present, in this article, a dynamical system constructed for
thermal convection in zero Prandtl number Boussinesq fluid, confined
between {\it stress-free} conducting flat boundaries, and subjected to
a slow rotation about the vertical axis. We then investigate
numerically the system to study the effect of coriolis force on the
onset of convection. We show that the convection sets in as
quasiperiodic waves at the onset of convective instability
for Taylor number $T$ above a critical value $T_c$, although
the {\it principle of exchange of stability} is valid according to
the linear theory~\cite{chandra} for these Taylor numbers.
The  generation of two independent frequencies is the result of
an interaction  between a stationary instability  and a
{\it self-tuned} wavy perturbations in presence of coriolis force.
This an example of a new {\it self-tuned} forward bifurcation.
For the values of Taylor number below $T_c$, the
convection sets in as {\it self-tuned} wavy instability as is the
case in absence of rotation. However, the model shows a possibility
transition from one wavy instability to another through a narrow
window of period-doubling instability.

We consider a thin layer of a Boussinesq fluid of infinite horizontal
extension subjected to a uniform adverse temperature gradient  $\beta$
across the fluid layer, and a rigid body rotation with an angular
velocity $\Omega $ about the vertical axis. The fluid is assumed to
have  uniform values of the kinematic viscosity $\nu$ and the thermal
difusitivity $\kappa$. The basic state is the conductive state with no
fluid motion in the rotating frame of reference. The convective flow,
in the limit of zero-Prandtl number, is then described  by
the following system of dimesionless hydrodynamic equations,
\begin{eqnarray}
\partial_t (\nabla^2 v_3) &=& \nabla^4 v_3 + R \nabla^2_H \theta
- \sqrt{T}\partial_z \omega_3 \nonumber\\
 & - &\hat{e}_3 {\bf \cdot  \nabla }\times
\left[({\bf \omega \cdot  \nabla }){\bf v } -
({\bf v \cdot  \nabla)} {\bf \omega}  \right], \label{vel}\\
\partial_t \omega_3 &=& \nabla^2 \omega_3 + \sqrt{T}\partial_z v_3 \nonumber\\
 & + & \left[({\bf \omega \cdot  \nabla }) v_3 -
({\bf v \cdot  \nabla}) \omega _3\right], \label{vort}\\
\nabla^2 \theta &=& - v_3, \label{temp}
\end{eqnarray}
where ${\bf v} (x,y,z,t) \equiv (v_1, v_2, v_3)$ is the velocity
field, $\theta (x,y,z,t)$ the deviation in temperature field from
steady conduction profile, ${\bf \omega} \equiv(\omega _1,
\omega_2, \omega_3 ) = {\bf  \nabla \times  v}$ the vorticity
field of the fluid. In the above,  length scales are made
dimensionless by  the thickness $d$ of the fluid layer, time by
the viscous time scale ${\frac{d^2}{\nu} }$, and the temperature
field by $(\beta d){\frac{\nu}{\kappa}}$. Rayleigh number
$R={\frac{\alpha g \beta d^4}{\nu \kappa} }$ and Taylor number
$T={\frac{4 d^4 \Omega^2}{\nu^2} }$ are the two dimensionless
external parameters. The unit vector $\hat{e}_3$ is directed
vertically upward. We impose periodic boundary conditions in
horizontal plane. This introduces two fundamental wave numbers
$k$ along x-axis and $q$ along y-axis. The stress-free boundary
conditions imply $\partial_z v_1$ $=$ $\partial_z v_2 $ $=$ $v_3$
$=0$ at $z=0, 1$. Thermally conducting horizontal boundaries
mean  $\theta = 0$ at $z = 0, 1$.

The hydrodynamical equations (\ref{vel}-\ref{temp}) are the same  as those
derived by Chandrasekhar~\cite{chandra}. We have nondimesionlized them and 
considered the case of zero $P$. We have also eleminited the pressure
term from Navier-Stokes equations by taking curl twice and using 
the incompressibility condition (${\bf \nabla \cdot v}=0$). 
The conclusions derived from the linearized version of the equations
remain unchanged even in the present case. Following the arguments of
Chandrasekhar~\cite{chandra}, one arrives at the conclusion that the
{\it principle of exchange of stability} is valid even in the limit
discussed here. The critical value of Rayleigh number now reads
$R_c (T) =\frac{\pi^4}{x}\left[(1+x)^3+\frac{T}{\pi^4}\right]$
and critical wave number $k_c (T)$ $=$ $\pi\sqrt{ (l_1 + l_2 - \frac{1}{2}) }$
now depend on Taylor number $T$~\cite{chandra}. In the above,
$l_{1,2}$ $=$
$\left[ \frac{1}{4} \left\{ \frac{1}{2}+\frac{T}{\pi^4}\pm
\sqrt{ \left( \frac{1}{2} + \frac{T}{\pi^4} \right) - \frac{1}{4}} \right\} \right]^{\frac{1}{3}}$
 and $x = k_c^2 / \pi^2$. In absence of rotation $k_c^0 \equiv k_c (T=0)$ $=$ $ \pi / \sqrt{2}$ and
$R_c^0 \equiv R_c (T=0)$ $=$ $ 27 \pi^4 /4$.

The 2D rolls are not exact solutions of nonlinear hydrodynamic
system with rotation as is the case in  zero-Prandtl number
convection in absence of rotation. Nevertheless, the growing 2D
rolls are  not saturated just above the onset of convective
instability. The  saturation  occurs only because of nonlinear
interaction of 2D rolls with 3D wavy perturbations.
To understand the nonlinear behavior close to
the convective instability, we construct a consistent 
minimal-mode model using Galerkin technique~\cite{mm}. 
We expand the vertical velocity $v_3$ and the vertical
vorticity $\omega_3$ in Fourier series compatible with the {\it
stress-free} boundary conditions and conducting thermal boundary
conditions. As the DNS, in absence of rotation, showed standing
patterns \cite{kft} instead of traveling patterns, we expect
similar behavior at least for small rotation rates. Therefore, we
expand the fields with real Fourier coefficients. This lead to
the following expansion for the  vertical velocity and
 the vertical vorticity in a minimum-mode model.
\begin{eqnarray}
v_3(x, y, z, t) &=& W_{101}(t)\cos{k_cx} \sin{\pi z} \nonumber\\
 &+& W_{111}(t)\cos{k_cx} \cos{qy} \sin{\pi z}\nonumber\\
&+& W_{{\bar 1}{\bar 1}1}(t)\sin{k_cx} \sin{qy} \sin{\pi z} \nonumber\\
&+& W_{012}(t) \cos{qy} \sin{2\pi z}  +  \ldots \\
\omega_3(x, y, z, t) &=& {\zeta}_{101}(t)\cos{k_cx} \cos{\pi z} \nonumber\\
&+& {\zeta}_{010}(t)\cos{qy} + {\zeta}_{{\bar 1}{\bar 1}0}(t) 
\sin{k_cx} \sin{qy}\nonumber\\ 
&+& {\zeta}_{111}(t)\cos{k_cx} \cos{qy} \cos{\pi z} \nonumber\\
&+& {\zeta}_{012}(t) \cos{qy} \cos{2\pi z} +
{\zeta}_{200}(t) \cos{2k_cx} \nonumber\\
&+& {\zeta}_{210}(t) \cos{2k_cx} \cos{qy}\nonumber\\
&+& {\zeta}_{{\bar 2}{\bar 1}0}(t)\sin{2k_cx} \sin{qy} + \ldots
\end{eqnarray}

The mode selection is quite systematic. As rotation couples
the vertical velocity and the vertical vorticity linearly, we
have selected the mode $\zeta_{101}$. The mode $\zeta_{010}$ is
essential to saturate zero Prandtl number convection via wavy
instability. All other modes appear through the nonlinear
interaction of these vorticity modes with the critical velocity mode
$W_{101}$. As the vorticity field is very crucial for saturation
in the limit of vanishing Prandtl number, all relevant second harmonics
are retained for vertical vorticity.  All relevant harmonics of the
vertical velocity field, consistent with the selection of the vertical
vorticity, are also retained.  Other higher order modes may be
required as Rayleigh number is raised further. As we
are interested to capture essential  nonlinear interaction between
competing instabilities just above the onset of convection, these
modes are essential. The solenoidal character of the velocity and
the vorticity fields yield horizontal components of the velocity
and the vorticity fields. The thermal fluctuation $\theta$ is captured
from Eq.~\ref{temp}. Projecting the hydrodynamic
equations~(\ref{vel} - \ref{temp}) on above modes, we arrive at
a twelve-dimensional dynamical system~\cite{sanjay}.

We now investigate the solutions of the dynamical system by
performing numerical integration of the model using standard
fourth order Runge-Kutta  as well as Bulirsh-Stoer schemes. By
choosing a value for $T$, we  set $k_c (T)$. We then choose a
value for $q$. We have tried with  different values of the
wavenumber $q$ of the perturbations and got qualitatively similar
results except when the ratio $q/k_c$ is close to unity. We
present here all the results for the case  $q/k_c(T)=0.4$.
Initial values for  all the twelve modes are chosen randomly, and
integration is done for a fixed value of Rayleigh number $R$. We
then repeat the process by increasing the value of $R$ in small
steps. We have also  tried various initial conditions. The
results of all the numerical integrations remain the same for the
same values of all the relevant parameters.  In absence of
rotation ($T=0$), only  six modes are excited. This  model then
reproduces the results of the model~\cite{kft} of zero P
convection without rotation. In presence of rotation all twelve
modes are excited as it should in a consistent model.

Figure 1 gives the stability boundaries of various
possible solutions, in the parameter space $R - T$, computed from
the model dynamical system.  The lowest line in this figure
shows linear dependence~\cite{chandra} of  $R_c$, the critical
Rayleigh number, on Taylor number $T$ for the onset of stationary
convection in zero Prandtl  number Boussinesq fluid. The onset of
overstability for the case  of vanishing P  for the Taylor
numbers considered here is much above, and is not shown in the
figure.

As Rayleigh number is raised above its critical value $R_c (T)$
for various values of Taylor number below $T=6.0$, conduction
state becomes unstable via stationary bifurcation. However,  2D
rolls with broken mirror symmetry~\cite{veronis} does not saturate 
until wavy perturbations interact with them. This saturates the 
growing rolls at finite amplitude even at the onset. The wavy  
perturbations are automatically generated when the amplitude of 
the 2D roll mode becomes large enough. The {\it self-tuned} 
waves consume the energy of 2D rolls and stop the unbounded  
growth of latter. This is precisely what happens when there is 
no rotation and 2D rolls have mirror symmetry.

The three rows of Fig. 2 show projections of the phase diagram,
starting from left, in $\zeta_{101}- W_{101}$,
$\zeta_{010}-W_{101}$, and $W_{111}-W_{101}$ planes for various
Rayleigh numbers (increasing downward) for fixed value of $T$ and
$q$.

As Rayleigh number is increased slowly, the solution changes from
one wavy solution to another through a thin regime showing period
doubling solutions (see the middle row of Fig. 2). The
first wavy solution $SW1$ has  2D mode $W_{101}$ with non-zero
mean as in the absence of rotation in zero P
convection~\cite{kft}, while the second wavy solution $SW2$ has
2D mode with zero mean. As Rayleigh number is increased, the
exchange of energy from 2D modes to waves increases. The larger
amplitude variation of vorticity modes is at the cost of energy of
2D rolls. This is well known feature in the case of oscillatory
instability.

As $T$ is raised further, the rotation facilitates  easily the
exchange of more energy from the 2D roll mode $W_{1010}$ to the
vertical vorticity mode $\zeta_{101}$ through linear coupling. We
observe an interesting behavior for $T>6$ (see Fig. 1).
The conduction state becomes unstable via stationary
instability~\cite{chandra} but the final state just above onset
is quasiperiodic waves~\cite{db}. Figure 3 shows the variation of
various modes with time. The amplitudes of all the modes begin
modulating at the same frequency. However, the ratio of the
frequency of wavy motion and that of amplitude modulation 
not an integer. The Fourier transform of these modes shows two
independent  frequencies. The frequency of amplitude modulation
is much smaller compared to that for wavy motion. The sharp
decrease of the amplitudes of higher order modes confirms the
fast convergence of the expansion. The model, therefore,
represents accurately the scenario close to the instability onset.
Figure 4 shows the projections of phase space
trajectories in various planes. It clearly describes the
quasiperiodicity of the convective flow. The trajectories are
confined in twelve dimensional torous in the phase space. The
quasiperiodic behavior originates due to the nonlinear
interaction among the 2D velocity mode $W_{101}$, the 2D vorticity mode
$\zeta_{101}$ excited by rotation, and the wavy vorticity mode
$\zeta_{010}$. Figure 5
reveals some interesting details of the time dependence of 
of convective patterns. The complex textures of the quasiperiodic
patterns are shown over a period of wavy motion, which is much faster
than the amplitude modulation. The halves of a period of wavy motion
are quite asymmetric. The textures of the pattern at different times 
are never the same due to quasiperiodicity. 


We have presented in this work a simple dynamical system, which
describes the phenomenon of thermal convection in rotating
Boussinesq fluid of zero Prandtl number very close to the onset.
For Taylor number above a critical value $T_c$, quasiperiodic
waves are observed at the instability onset. For very values of
Taylor number below $T_c$, the coriolis force causes one wavy
instability to another through period doubling instability. 
We have shown that convection might be possible as quasiperiodic
waves, even if the {\it principle of exchange of stability} is valid
according to linearized hydrodynamical system. The saturation to
quasi-periodic convectie state is {\it self-tuned} purely due to 
the nonlinear effects. This is an example of new {\it self-tuned} 
bifurcation scenario from the conduction state to unsteady 
convective state. The model presented would also be useful to study 
an interesting possibility of transition from a state of rest to 
quasi-periodic chaos~\cite{rt} at the primary instability.

Acknowledgements: This work was sponsored by DST, Govt. of India under
its project ``Pattern-forming instability and interface waves".\\

\newpage
\begin{center}
{\bf Figure Caption}
\end{center}

Fig. 1: Stability boundaries in the parameter space $R-T$ 
  just above the onset of convection. The conduction state 
  is stable below the lowest straight line, which shows 
  critical value  $R_c$ of Rayleigh number as a function of 
  Taylor number $T$ ($q/K_c=0.4$). The region marked $QP$  
  shows quasiperiodic  waves. The regions denoted by $SW1$
  and $SW2$ show two regimes of wavy solutions separated by 
  a thin region where period-doubling is  observed.\\

Fig. 2: Phase portrait for various values of Rayleigh number 
  $R$ ($T=2.0$ and $q/k_c=0.4$). The top row, starting from
 left, shows the plots of modes $\zeta_{101}$, $\zeta_{010}$, and
 $W_{111}$ with respect to $W_{101}$ for $R=663.0$. The middle and
 the  bottom rows show the same plots for $R=664.95$ and $R=668$
 respectively. The top and the bottom rows two different wavy regimes
 $SW1$ and $SW2$ respectively, while the middle row represents period
 doubling corresponding to the narrow regime between $SW1$ and $SW2$
 in Figure 1.\\

Fig. 3: Time variation of all the modes for $T=10.0$, $q/k_c=0.4$,
   and $R=678.0$ long after all transients are died out . The critical
   Rayleigh number $R_c(T=10)=677.0768$. Starting from left, the
   top row shows variation of $W_{101}$, $W_{{\bar 1}{\bar 1}1}$,
   $W_{111}$, and $W_{012}$ with time. The middle row shows variation
   of $\zeta_{101}$, $\zeta_{010}$, $\zeta_{111}$, $\zeta_{200}$, and
  the bottom row shows variation of $\zeta_{012}$, $\zeta_{210}$,
   $\zeta_{{\bar 2}{\bar 1}0}$, and $\zeta_{{\bar 1}{\bar 1}0}$.\\

Fig. 4: Phase space portraits showing quasiperiodic motion for
   $T=10.0$, $q/k_c=0.4$, and $R=677.08$. Starting clockwise from
   the left top, they show the projections of  of the phase space in
   $W_{101}-W_{{\bar 1}{\bar 1}1}$, $\zeta_{101}-W_{101}$,
   $\zeta_{{\bar 2}{\bar 1}0}-W_{101}$, and $W_{012}- W_{101}$ planes.
   Some modes like $\zeta_{{\bar 2}{\bar 1}0}$ and $W_{012}$ have faster
   time period (i.e., period of wavy oscillation) double that of
   $W_{101}$.\\

Fig. 5: Contour plots ($T=7.0$, $q/k_c=0.4$, $R=671.30$) at $z=0.25$.
 Various textures of the pattern is shown over a period of
 faster time scale $t_0$ at the equal time interval of $t_0/8$. The
 sequence of time evolution of the pattern-texture is shown from left
 to right in each row starting from the top row.

\end{document}